\documentclass[twocolumn,showpacs,preprintnumbers,amsmath,amssymb]{revtex4}

\usepackage{graphicx}
\usepackage{amssymb}
\usepackage{dcolumn}

\begin{document}
\title{Coevolutionary dynamics on scale-free networks}

\author{Sungmin Lee}
\author{Yup Kim}
\email{ykim@khu.ac.kr  } \affiliation{Department of Physics and
Research Institute for Basic Sciences, Kyung Hee University, Seoul
130-701, Korea}

\begin{abstract}
We investigate Bak-Sneppen coevolution models on scale-free
networks with various degree exponents $\gamma$ including random
networks. For $\gamma >3$, the critical fitness value $f_c$
approaches to a nonzero finite value in the limit $N \rightarrow
\infty$, whereas $f_c$ approaches to zero as $2<\gamma \le 3$.
These results are explained by showing analytically $f_c(N) \simeq
A/<(k+1)^2>_N$ on the networks with size $N$. The avalanche size
distribution $P(s)$ shows the normal power-law behavior for
$\gamma >3$. In contrast, $P(s)$ for $2 <\gamma \le 3$ has two
power-law regimes. One is a short regime for small $s$ with a
large exponent $\tau_1$ and the other is a long regime for large
$s$ with a small exponent $\tau_2$ ($\tau_1 > \tau_2$). The origin
of the two power-regimes is explained by the dynamics on an
artificially-made star-linked network.
\end{abstract}
\maketitle

Bak and Sneppen (BS) \cite{BS} has introduced an excellent model
to explain the evolution of bio-species which exhibits the {\it
punctuated equilibrium} behavior \cite{KJ}. BS model has two
important features, coevolution of the interacting species and the
intermittent bursts of activity separating relatively long periods
of the stasis. In BS model the ecosystem evolves into a
self-organized criticality with avalanches of mutations occurring
all scales. Aside from its importance for the evolution BS model
has been also shown to have rich scaling behaviors \cite{PMB}.

Since BS model was suggested, the model has been extensively
studied on regular lattices or networks \cite{PMB}. However, many
important bio-systems have been elucidated to form nontrivial
networks by the recently developed network theories \cite{ABDM}.
Important examples are metabolic network, cellular network, and
protein network \cite{JTAOB,JMBO,WM,CGA}. Especially the important
bio-networks are scale-free networks (SFNs) \cite{ABDM}, in which
the degree distribution $p(k)$ satisfies a power law $p(k) \sim
k^{-\gamma}$  \cite{ABDM}. Thus it is important to study the BS
dynamics on SFNs or to find out how the base structure of
interacting biological elements (cells, proteins, or species)
affects the evolutionary change or dynamics of the given
bio-system. Until now BS models on the nontrivial networks were
not investigated extensively. Christensen et al. \cite{CDKS} have
studied BS model on random networks (RNs). Kulkani et al.
\cite{KAS} studied BS model on small-world networks. Slania and
Kotrla \cite{SK} studied the forward avalanches of a sort of
extremal dynamics with evolving networks. Moreno and Vazquez
\cite{MV} studied BS model only on a SFN with $\gamma=3$.

In this letter, we will study BS models on SFNs in complete and
comprehensive ways. One of the main purposes of this study is to
find which structure of interacting species is the most stable
network or most close to mutation-free network under the
coevolationary change with interacting species. As is well-known,
SFNs with the degree exponent $2<\gamma \le 3$ are physically much
different from those with $\gamma > 3$ \cite{ABDM}. We study BS
models not only on SFNs with $2<\gamma \le 3$ but also on SFNs
with $\gamma > 3$ including random networks (or SFN with
$\gamma=\infty$). As we shall see, two important results are found
in this study. First, the critical fitness value $f_c$ of BS
models for $\gamma \le 3$ is shown to have the limiting behavior
$f_c(N) \rightarrow 0$ when the number of nodes $N$ of the network
goes to infinity. In contrast, $f_c$ approaches finite nonzero
value as $N \rightarrow \infty$ for $\gamma>3$. Furthermore, $f_c
(N)$ on SFNs with finite $N$ is shown to satisfy the relation
$f_c(N) \simeq {{const.} \over {<(k+1)^2>_N}}$, which is also
directly supported by simulation. Second, for $2<\gamma \leq 3$
the distribution of avalanches is shown to have two power-law
regimes. To find the origin of this anomalous behavior of
avalanches we also study BS models on an artificially-made
star-linked network and find the similar two power-law regimes.

We now explain the model treated in this letter. All the models
are defined on a graph $Gr=\{N,K\}$, where $N$ is the number of
nodes and $K$ is the number of degrees with the average degree
$<k>=2K/N$. Initially, a random fitness value $f_i \in [0,1]$ is
assigned to each node $i=1,...,N$. At each time step, the system
is updated by the following two rules: (I) first assign new
fitness value to the node with the smallest fitness value
$f_{min}$. (II) Second assign new fitness values to the nodes
which are directly connected to the node with $f_{min}$. We use
SFNs with the various degree exponent $\gamma$ as $Gr=\{N,K\}$. To
generate SFNs, we use the static model \cite{SNUGroup} instead of
preferential attachment algorithm \cite{ABDM}.

\begin{figure}
\includegraphics[width=7cm]{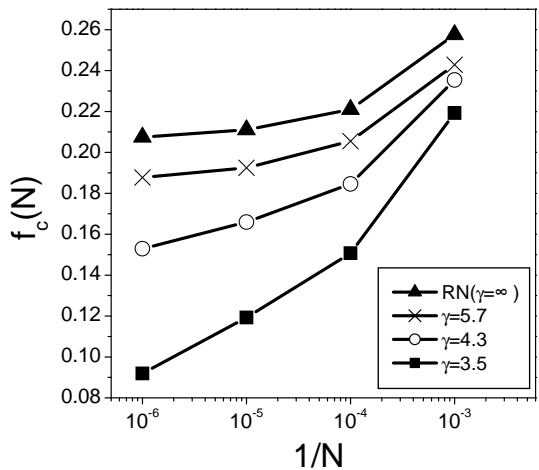}
\caption{Semi-log plot of the threshold $f_c (N)$ versus $1/N$ on
RN and on SFNs with $\gamma=5.7$, $4.3$, and $3.5$. Used networks
sizes for each network are $N=10^3$, $10^4$, $10^5$, and $10^6$.
The solid lines between data points are obtained by simple linear
interpolations.}
\end{figure}

To understand the dependence of the critical fitness value
$f_c(N)$ on $\gamma$, we generate SFNs with $\gamma= \infty, 5.7
\sim 2.15$. To exclude the effects of finite percolation clusters
\cite{CDKS} and to see the effect of network structure itself, all
the networks are made to have average degree $<k>=4$. To
understand the dependence on number of nodes $N$, the networks
with the sizes $N=10^3 \sim 10^6$ are generated for each $\gamma$.
To determine the critical fitness value $f_c (N)$, we consider
$f_{min}$ as a function of the total number of updates $s$
\cite{PMB}. Initially, $f_{min} (s=0)$ is the gap $G(0)$, where
$G(s)$ is the maximum of all $f_{min} (s')$ for $0 \leq s' \leq s$
\cite{PMB}. When $G(s)$ jumps to a new higher value, there are no
nodes in the system with $f_i (s) < G(s)$. Thus $\lim_{s
\rightarrow \infty} G_N (s)=f_c (N)$.

We measure $f_c(N)$ on the various SFNs. Fig. 1 shows the plot of
$f_c (N)$ versus $1/N$ for SFNs with various $\gamma$. The values
of critical fitness $f_c (N \rightarrow \infty)$ evaluated from
data in Fig. 1 are $0.21(1)$, $0.19(1)$, $0.15(1)$, and $0.09(1)$
for $\gamma= \infty$, $5.7$, $4.3$, and $3.5$. The results in Fig.
1 mean that for $\gamma> 3$, $f_c (N \rightarrow \infty)
\rightarrow const.(>0)$.

Fig. 2 shows the plot of $f_c (N)$ versus $1/N$ for $2<\gamma \le
3$. For $\gamma=3$, $f_c (N)$ nicely satisfies the relation, $f_c
(N) \sim 1/ \ln N$ \cite{MV}. For $2 < \gamma <3$, $f_c (N) $'s
seem to follow a power-law $f_c (N) \sim N^{-\eta}$ and approach
to zero as $N$ goes to $\infty$. In contrast to the results in
Fig. 1, $f_c \rightarrow 0$ for $2<\gamma \le 3$.

In the RN, every pair of nodes are randomly connected and the
degree distribution is a Poisson distribution \cite{ABDM,CDKS}. So
the BS model on RN \cite{CDKS} is a good realization of the
mean-field-type random neighbor model. In the random neighbor
model, the fitness values of the randomly selected $(m-1)$ nodes
as well as the node with $f_{min}$ are updated and $f_c= 1/m$
\cite{FSB}. The result $f_c(\infty)=0.21(1)$ on RN is very close
to $ {\frac{1}{<k>+1}} = {\frac{1}{5}}$, which is expected one
from the random neighbor model by setting $<k>+1=m$ \cite{CDKS}.
In the steady state of BS model, the probability measure $P(f <
f_c )$ is $0$. Suppose the case that the number of updates for
each step is fixed as $m$, as in the random neighbor model. To
sustain the steady state in the case, at most one new fitness
value should be less than $f_c$ and the other $m-1$ new values
should be larger than $f_c$ \cite{FSB}. Therefore we can easily
see $ m f_c = 1$ or $f_c = 1/m$.

On a network the number of updats depends on the degree of the
node with $f_{min}$ and the probability which a node with degree
$k$ is connected to the node with $f_{min}$ should be proportional
to $k$. For an updating step the probability that a node with
degree $k$ is updated is proportional to $k+1$, because the node
itself can be the node with $f_{min}$. Therefore, after an
arbitrary update, the probability $P_{min} (k)$ of a node with
degree $k$ being the node with $f_{min}$ is proportional to $k+1$.
This means that $P_{min} (k)$ in the steady state should be
proportional to $k+1$, or $P_{min} (k) = \frac{(k+1)p(k)}{\sum_k
(k+1)p(k)} = \frac{1}{<k>+1}(k+1)p(k)$. The average number
$N_{update}$ of the nodes updated for one updating process is
therefore
\begin{eqnarray}
N_{update} = \sum_k (k+1) P_{min} (k) = \frac{\sum_k(k+1)^2
p(k)}{<k>+1}
\end{eqnarray}
and thus $f_c$ is
\begin{eqnarray}
f_c = \frac{1}{N_{update}} = \frac{<k>+1}{\sum_k (k+1)^2 p(k)}
=\frac{<k>+1}{<(k+1)^2>}. \label{FC}
\end{eqnarray}
When the number of updates is fixed as $m$, Eq. (\ref{FC})
reproduces the mean-field result $f_c = 1/m$. In SFNs with $p(k)
\simeq k^{-\gamma}$, Eq. (\ref{FC}) becomes
\begin{eqnarray}
f_c \simeq \left \{ \begin{array}{ll} finite, & \gamma >3 \\
\frac{A}{<k^2>} = \frac{A}{\int k^{2-\gamma} dk} , &   2 < \gamma
\le 3.  \end{array} \right. \label{FCS}
\end{eqnarray}
Eq. (\ref{FCS}) explains the results in Figs. 1 and 2 including
the result $f_c \simeq \frac{1}{\ln N}$ for $\gamma=3$. For
$2<\gamma < 3$, measured $f_c (N)$ is fitted to the relation
$f_c(N) = A /<k^2>_N$, where $A$ is constant and $<k^2>_N$ is
$<k^2>$ for the network with the size $N$. The fitted lines in
Fig. 2 show that the relation $f_c(N) = A /<k^2>_N$ holds well and
directly supports Eq. (\ref{FCS}).

\begin{figure}[b]
\includegraphics[width=7cm]{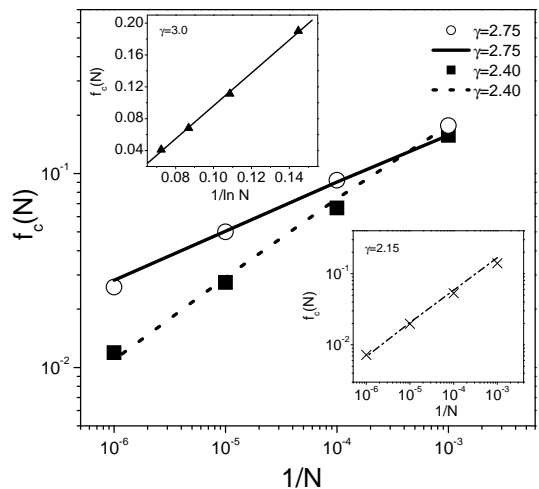}
\caption[0]{Log-log plot of $f_c (N)$ and $A/<k(N)^2>_N$ versus
$1/N$ on SFNs with $\gamma=2.75$, $2.40$, and $2.15$. Symbols are
for $f_c (N)$ and the lines are for $A/<k(N)^2>_N$, where $A$ is a
constant. The top inset shows the plot of $f_c (N)$ versus $1/\ln
N$ for $\gamma=3.0$.}
\end{figure}

An avalanche in Bak-Sneppen model is defined as the sequential
step $s$ for which the minimal site has a fitness value smaller
than given $f_o$ \cite{PMB}. For each network, we choose $f_o$ to
satisfy $(f_c (N) - f_o)/f_c (N)= 0.05$. The probability
distribution $P(s)$ of avalanche size $s$ on the networks with the
size $N=10^6$ are shown in Fig. 3 and Fig. 4. All the data in
Figs. 3 and 4 are taken in the steady-states.

\begin{figure}
\includegraphics[width=7cm]{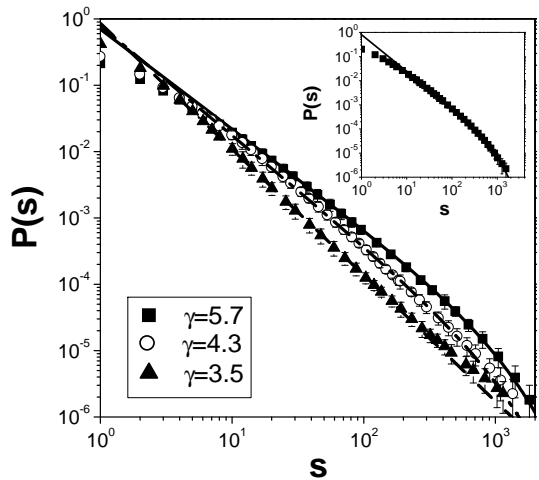}
\caption[0]{Log-log plot of the avalanche size distribution $P(s)$
on SFNs with $\gamma=5.7$, $\gamma=4.3$, $\gamma=3.5$ and on RN
(Inset). The curves for $\gamma=5.7$, $\gamma=4.3$ and RN denote
the fits of the form $P(s)=As^{- \tau} exp(-s/s_c )$ to the data.
Obtained exponents are $\tau = 1.5$ for both $\gamma=5.7$ and RN,
and $\tau = 1.65$ for $\gamma=4.5$. The line for $\gamma=3.5$
denotes the fit of the form  $P(s)=As^{-\tau}(\tau=1.65)$ without
cutoff.}
\end{figure}

As is shown in Fig. 3, $P(s)$ in SFNs with $\gamma > 3$ including
RN satisfy the normal power-law behavior with an exponential
cutoff as $P(s)= As^{- \tau} \exp(-s/s_c )$. The curves in Fig. 3
represent the fitted curves to data for $P(s)$. From those
fittings the obtained values for $\tau$ are 1.5 for RN and
$\gamma= 5.7$, and  1.65 for $\gamma=4.3$. The result for RN and
SFN with $\gamma = 5.7$ is expected from the random neighbor model
\cite{FSB}. As $\gamma$ decreases to 4.0 or so $\tau$ increases to
$1.65$. For $\gamma=3.5$, however, the best fitting function is
$P(s)=Bs^{-\tau}$ with $\tau=1.65$ and we cannot find the
cut-off-dependent behavior within our data. Instead, it is even
observed that tails of measured data for $\gamma=3.5$ around
$s=10^3$ seem to deviate from the fitting function $P(s)=Bs^{-
\tau}$ and are lager than values estimated from the best fitting
function. This rather anomalous tail behavior of $P(s)$ for
$\gamma=3.5$ should be the signal of the anomalous behavior of
$P(s)$ for $2<\gamma \le 3$.

In contrast to the simple power-law behavior for $\gamma >3$,
anomalous behavior for $P(s)$ shows up for $2<\gamma \le 3$ (Fig.
4). We can see two power-law regimes clearly for $P(s)$ in Fig. 4.
Initially the avalanche size distribution follows $P(s) \simeq
s^{-\tau_1}$ about 1 decade or so. After this short initial
power-law regime, the long second power-law regime appears as
$P(s) \simeq s^{-\tau_2}$, where $\tau_1 > \tau_2$. The measured
exponents $\tau_1$, $\tau_2$ are summarized in Table I.

\begin{table}
\caption{\label{tab:table1}Two power-law exponents, $\tau_1$ and
$\tau_2$ for SFNs with $\gamma \le 3$.}
\begin{ruledtabular}
\begin{tabular}{lcr}
$\gamma$ &$\tau_1$ & $\tau_2$\\
\hline
3.0 & 2.09 & 1.59\\
2.75 & 2.22 & 1.47\\
2.4 & 2.27 & 1.32\\
2.15 & 2.30 & 1.20\\
\end{tabular}
\end{ruledtabular}
\end{table}

Compared to the behavior of the avalanche size distribution for
$\gamma > 3$, this anomalous behavior of $P(s)$ is very peculiar.
In the steady state, it is expected that the node with $f_{min}$
(the minimal node) is most frequently found among the last updated
nodes \cite{FSB} and then the minimal node locally performs a
random walk. However, there can be longer jumps of any length with
a very low probability. If this kind of a {\it jumpy random walk}
is the motion of the minimal node, then a subnetwork consists of a
{\it hub node} ({\it center node}) and many {\it slave nodes}
directly linked to the hub should be important to decide the
behavior of $P(s)$. Due to the {\it jumpy random walk} behavior,
the more slave nodes the hub node has, the longer stay of the
minimal node or the longer avalanche exists at the given
subnetwork. This effect explains the second power-law regime with
the exponent $\tau_2$ in Fig. 4, because $<k^2>$ diverges for
$2<\gamma \le 3$, and so the subnetwork of a hub node and many
slave nodes should be the main substructure in SFNs with $2<\gamma
\le 3$. Evidently, the jumpy steps of the {\it jumpy random walk}
make the shorter avalanches possible and this effect explains the
first power-law regime with the exponent $\tau_1$.

\begin{figure}[b]
\includegraphics[width=7cm]{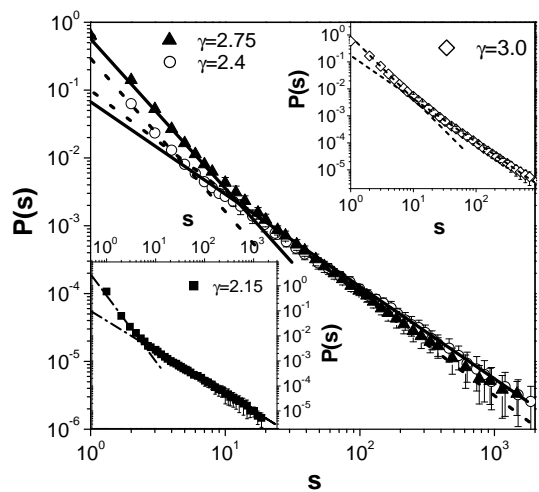}
\caption[0]{Log-log plot of $P(s)$ on SFNs with $\gamma=3$ (top
inset), $2.75$, $2.4$, and $2.15$. Two crossing lines for each
data sets denote the two power-law regimes, $P(s)=As^{- \tau_1 }$
and $P(S)=Bs^{- \tau_2 }$. Obtained exponents, $\tau_1$ and
$\tau_2$, are shown in Table I.}
\end{figure}

To support the qualitative explanation of the two power-law
regimes, we consider an artificially-made star-linked network
shown in Fig. 5. In the star-linked network, a main subnetwork
consists of a center (star) node and many dangling slave nodes
linked directly to the star node. Then the center nodes are linked
hierarchically to one after another as sketched in Fig. 5(a). We
make a star-linked network in which there are $25$ base
subnetworks with $500$, $480$, $...$, and $20$ slave nodes,
respectively. In this network, we perform BS dynamics and find
$f_c = 0.123$. $P(s)$ is also measured on the star-linked network
and is shown in Fig. 5(b). We find the very two power-law regimes
with the exponents $\tau_1=3.7$ and $\tau_2=1.27$. The plateau
between two power-regimes in the data of $P(s)$ in Fig. 5(b) is
probably from the discrete distribution of the number of slave
nodes.

In conclusion, we study BS models on SFNs with various $\gamma$.
For $\gamma >3$, $f_c$ approaches to a nonzero value in the limit
$N \rightarrow \infty$ and $P(s)$ shows normal power-law behavior
with $\tau \ge 1.5$. For $\gamma \le 3$, $f_c$ approaches to zero
as $f_c(N) \simeq A/<K^2>_N$ and $P(s)$ has two power-law regimes.
The origin of the two power-regimes are explained by the dynamics
on a star-linked network.

\begin{figure}
\includegraphics[width=7cm]{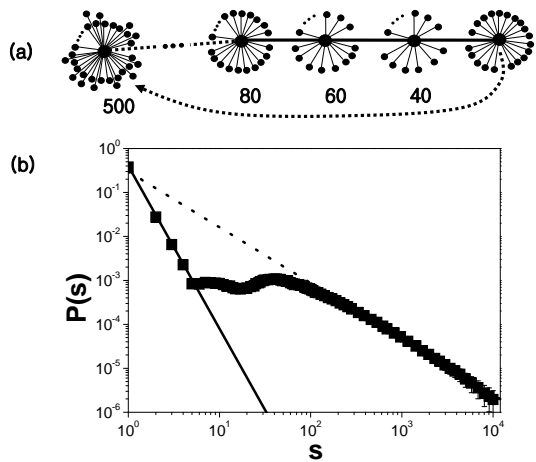}
\caption{(a) Schematic diagram of a star-linked network which
consists of 25 subnetworks with 500, 480, ..., and 20 dangling
slave node. (b) Plot of $P(s)$ on the star-linked network
structure. Two power-law regimes with $P(s)=As^{- \tau_1 }(\tau_1
= 3.7)$ and $P(s)=Bs^{- \tau_2 } (\tau_2 = 1.27)$ are clearly
shown by the lines.}
\end{figure}

In Ref. \cite{MV}, BS dynamics only on a SFN with $\gamma=3$ was
studied and the only meaningful numerical result was to show
$f_c(N) \simeq 1/\ln N$. Ref. \cite{MV} suggested a relation
similar to Eq. (\ref{FC}) from a rate equation which was obtained
by a naive and immature analogy of BS dynamics to the epidemic
dynamics on SFNs \cite{PV}. However the rate equation should never
be the exact one. Even the exact rate equation for the simple
random neighbor model \cite{FSB} is much more complex than that of
Ref. \cite{MV} or the epidemic dynamics. The correct rate equation
for BS dynamics on SFNs must be derived by considering all the
terms of the rate equation in Ref. \cite{FSB} and the base network
structure simultaneously and correctly. The derivation of the
correct rate equation should be a subject for the future study. In
Ref. \cite{MV} they argued $P(s)$ for $\gamma=3$ satisfies a
simple power-law with $\tau \simeq 1.55$. By the brute-forced fit
of the relation $P(s)\simeq s^{-\tau}$ to our data in Fig. 4, we
also obtain $\tau \simeq 1.6$ for $\gamma=3$. However, this blind
application of the simple power law should be wrong and there
should exist the two-power law regimes even for $\gamma=3$. One
can easily identify the two power-law regimes in the $P(s)$ data
of Ref. \cite{MV} rather clearly although the tail parts of their
data are qualitatively poor and show large fluctuations.

The occurrence of two power-law regimes for $P(s)$ was also found
in BS dynamics on small-world networks \cite{KAS} and in an
extremal dynamics with evolving networks \cite{SK}. However the
origins of the two power-law regimes were completely different
from ours. The origin in the small-world networks was argued to be
the long range connectivity of the networks \cite{KAS}. The
extremal dynamics with evolving random networks \cite{SK} changes
the network structure and is not exactly the same as BS dynamics.
Furthermore the evolving network develop many disconnected
clusters. In the model \cite{SK} the forward avalanches
are mainly measured. The forward avalanches\cite{SK} should be affected
by the dynamical aggregate and
splitting of subnetworks by the extremal dynamics, which should be
the origin of the two power-law regimes. In contrast our
avalanches of BS dynamics is measured on a fully-connected static
scale-free network and should not be directly comparable to
the avalanches on dynamically varying networks.

Authors would like to thank Prof. H. Jeong for valuable
suggestions. This work is supported by Korea Research Foundation
Grant No. KRF-2004-015-C00185.

\end{document}